\documentclass[a4paper,
               keeplastbox,   
               ]{jacow}

\begin{document}

\title{Drift Observations and Mitigation in LCLS-II RF\thanks{This work was supported by the LCLS-II Project and the U.S. Department of Energy, Contract DE-AC02-76SF00515}}

\author{L. Doolittle\thanks{lrdoolittle@lbl.gov},  S.D. Murthy, LBNL, Berkeley, CA 94720, USA\\
A. Benwell, D. Chabot, J. Chen, B. Hong, S. Hoobler, J. Nelson, C. Xu, SLAC, Stanford, CA 94309, USA }

\maketitle

\abstract
The LCLS-II RF system physically spans 700\thinspace m and has strict requirements --- on the order of 20\thinspace fs --- on the phase stability of the accelerating RF fields in its SRF linac.  While each LLRF rack is crudely temperature-stabilized, the weather inside the service building as a whole is usually compared to a tin shack in the California sun. A phase-averaging reference line is the primary system deployed in support of the phase stability goals. There are other, secondary subsystems (SEL phase offset, and determination of cavity detuning) that are also sensitive to RF phase drift. We present measurements of phase shifts observed in the overall RF system, and how diagnostics are able to sense and correct for them during beam operations.

\section{INTRODUCTION}

LCLS-II SRF cavity field control~\cite{lcls2} is performed with a digital
Delayen-style~\cite{delayen} Self-Excited-Loop (SEL).
Cavity phase is measured relative to a Phase Reference Line (PRL).
Resonance control is effected by piezo actuators (and stepper motors)
based on state-space interpretation of forward and cavity waveforms.

\begin{figure}[hb]
    \centering
    \includegraphics[scale=0.09]{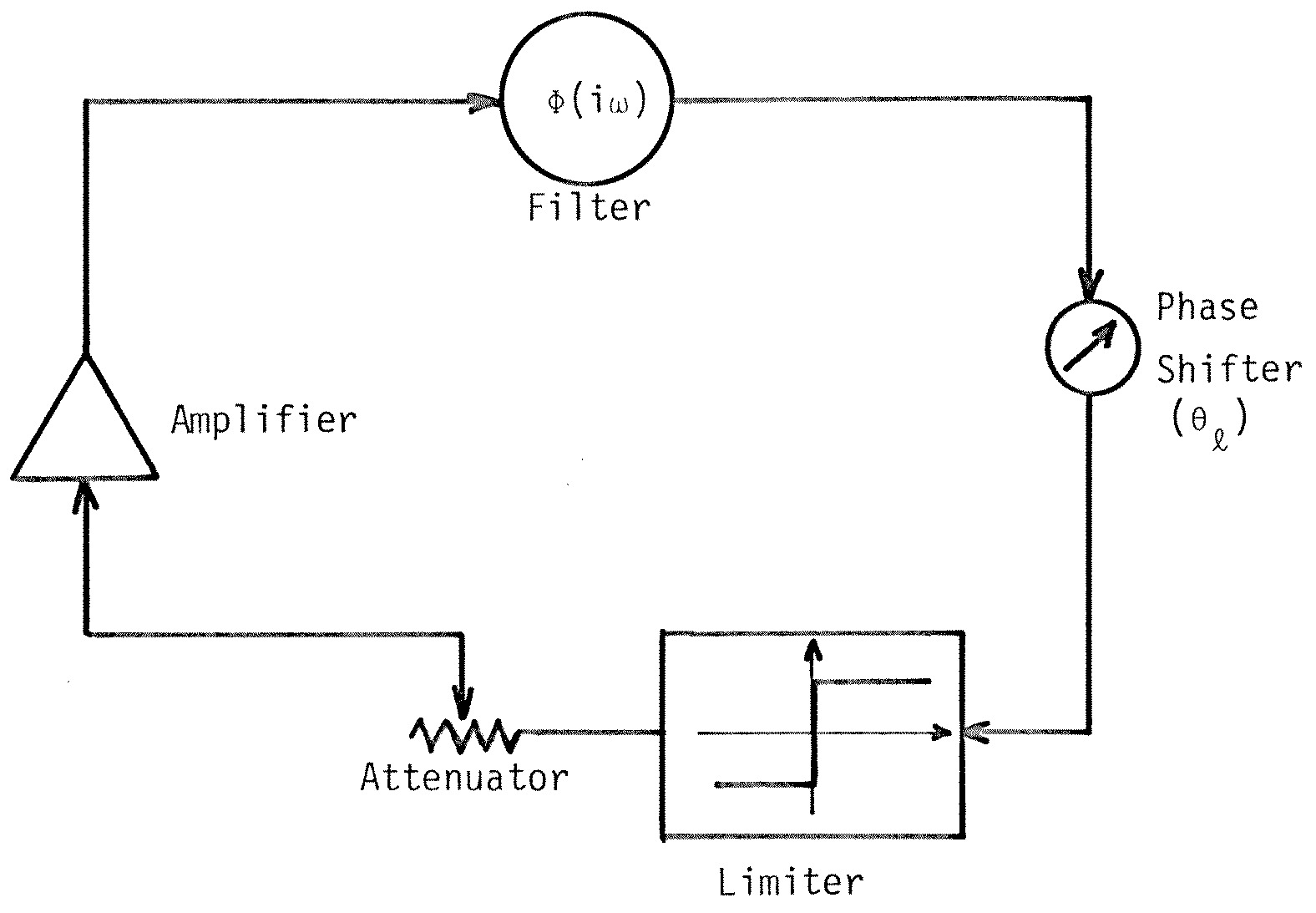}
    \caption{Analog SEL from Delayen.}
\end{figure}
\begin{figure}[hb]
    \centering
    \includegraphics[scale=0.35]{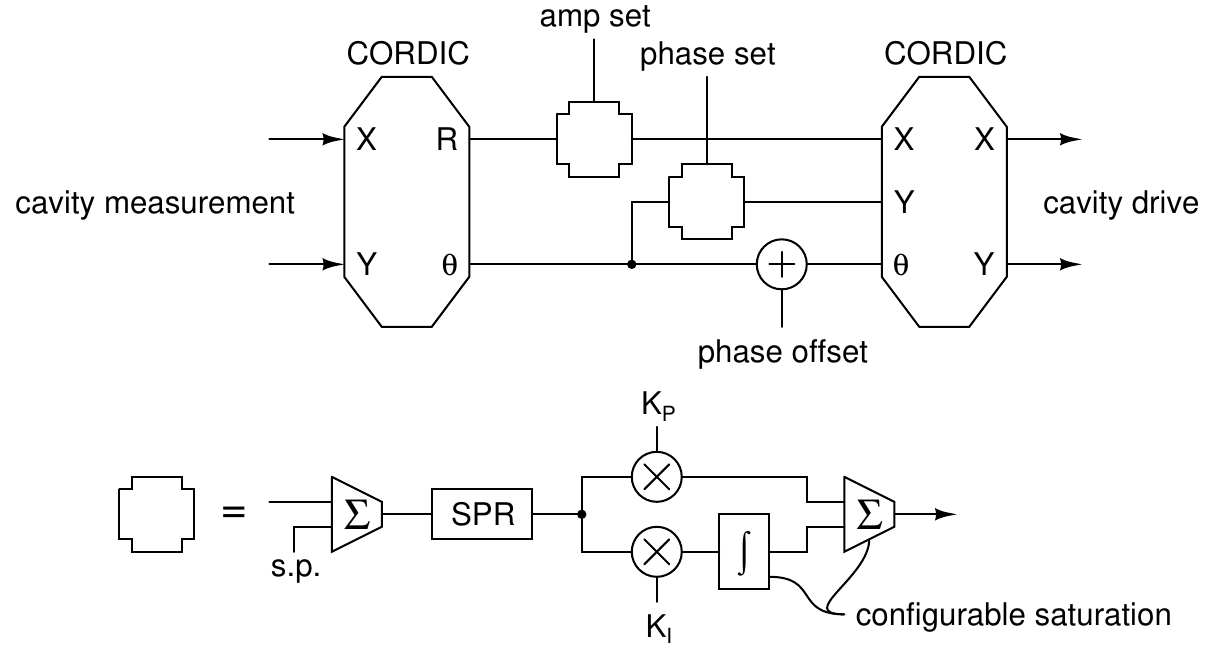}
    \caption{Digital SEL with CORDIC and PI loops.}
\end{figure}

The dominant topic of this paper is deployment of SRF cavity control
with feedback loops ``at scale.''  The entire machine consists of
280 $\times$ 1.3\thinspace GHz SRF cavities
and 16 $\times$ 3.9\thinspace GHz SRF cavities, not considering the
gun and bunchers that use copper cavities.  Each SRF cavity
needs its own field control loop (really two loops in one,
considering I \& Q or magnitude \& phase) and resonance control loop.
Many more feedback loops are used in the Phase-averaging Reference Line (PRL)
and the tracking of that signal in Low Level Radio Frequency (LLRF) system.

Both the field control and resonance control loops are subject to errors
creeping in when RF cables change electrical length due to temperature variations,
leading to observable system phase shifts.
We deploy additional self-tuning feedback loops to mitigate those effects.
Those loops depend on the presence of microphonics (time-varying cavity resonance frequency)
in the SRF cavities.

Finally, entire linacs are wrapped in beam diagnostics that can stabilize
energy gain and chirp of the ensemble.  To the LLRF system, this is the
meaningful out-of-loop error indicator.

\section{CAPTURE OF PRL SIGNAL}

A bi-directional phase-averaging reference line is deployed in LCLS-II~\cite{prl1}.
A single cavity control rack consists of three chassis: a Precision Receiver Chassis (PRC) and two RF Stations (RFS).
Both directions of the traveling-wave phase reference in the long
coax are down-mixed and digitized in the PRC, and the final phase averaging
is done in FPGA DSP~\cite{prl2}.

\begin{figure}[hbt]
    \centering
    \includegraphics[width=0.4\textwidth]{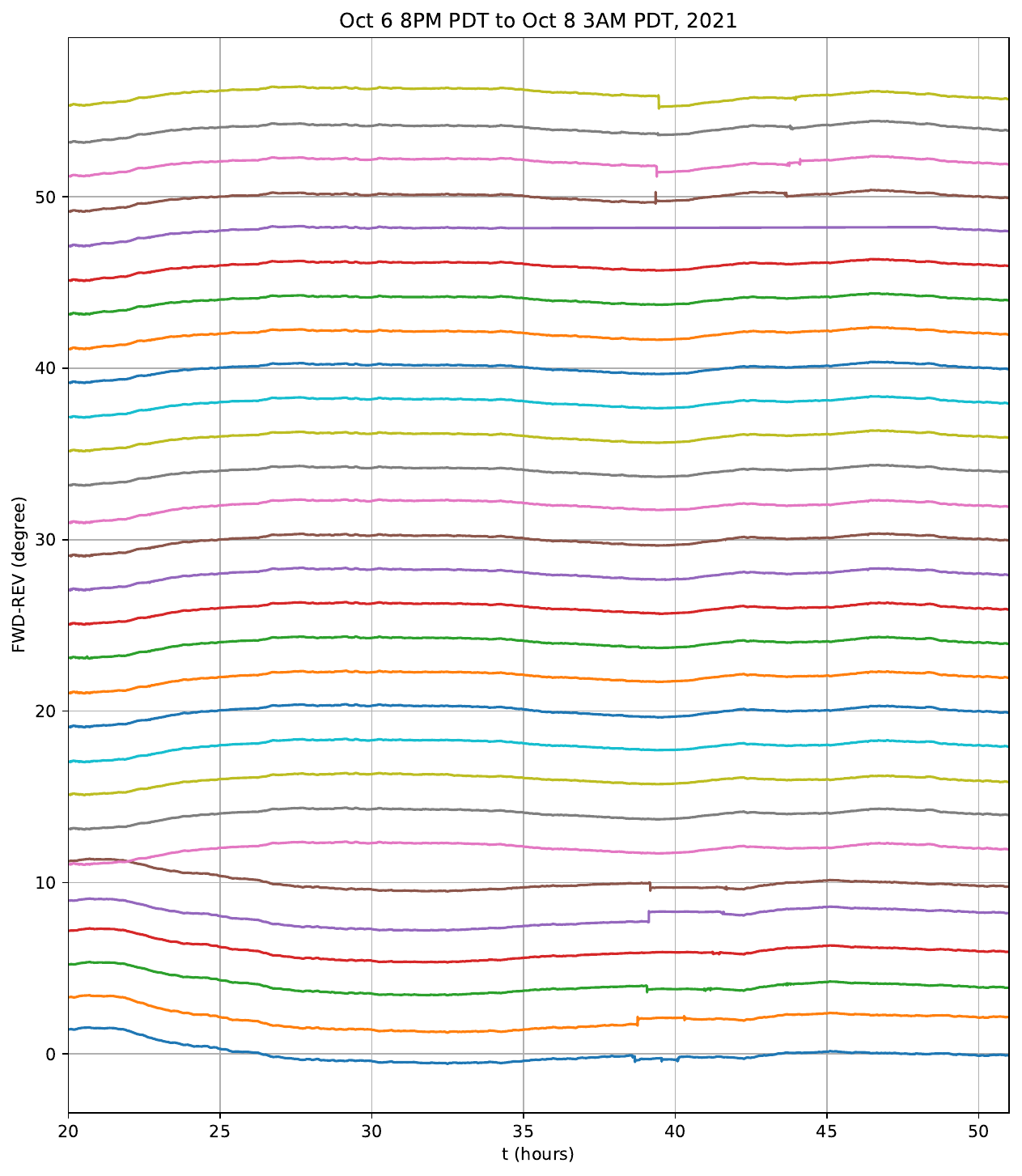}
    \caption{Observed PRL drift (Fwd-Rev).}
    \label{prl_fwd_rev}
\end{figure}

The difference between PRL forward and reverse phase measurements in
a PRC generally represents the phase reference cable
itself changing length due to temperature changes.
Figure~\ref{prl_fwd_rev} shows this effect for some (L0, L1 and L2) cryomodules
over two days, with a median span of 1.3$^\circ$.
This cable is installed underground in the accelerator tunnel,
where it is isolated from harsh day-night temperature swings.

\begin{figure}[hbt]
    \centering
    \includegraphics[width=0.4\textwidth]{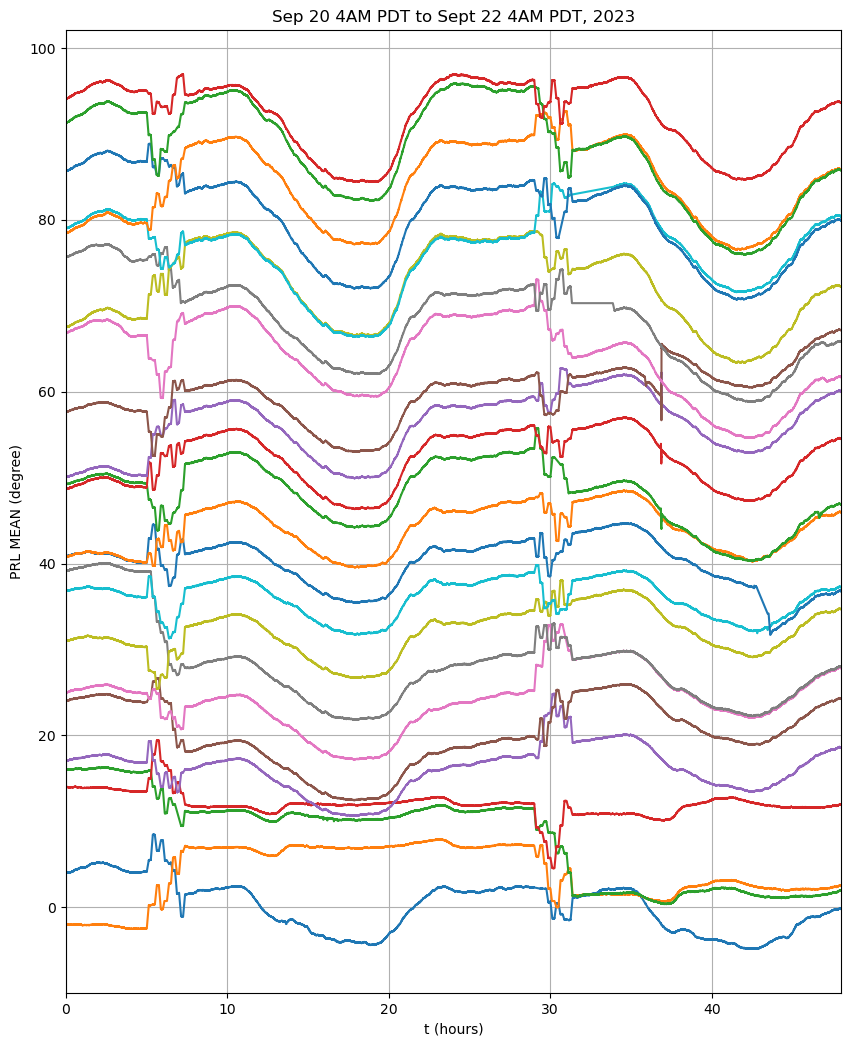}
    \caption{Observed LO drift.}
    \label{prl_zero_drift}
\end{figure}

The mean of those forward and reverse phase measurements
generally represents the Local Oscillator (LO) distribution cable changing length.
That physical 150\thinspace m (in L2) LO cable is in
the gallery, which is not temperature controlled, and subject to greater temperature induced phase changes as shown in figure~\ref{prl_zero_drift}.
The observed median day-night span is typically 14.8$^\circ$.

A cavity control rack (used to control 4 SRF cavities, servicing half a cryomodule)
can run mostly OK without input from the \hbox{PRL}.
That mode of operation will operate cavities at the right frequency, but with unknown
absolute phase (or even phase relative to its neighbor).
That's still perfectly useful for cavity commissioning and testing,
and in fact much of this project's firmware and software was initially
developed and tested in this context.
Such operations are clearly not useful for running beam.
Use of the PRL eliminates sensitivity to LO distribution phase drifts
and chassis-level divider state, so cavities
don't need to be re-phased when chassis are restarted.

\section{SEL PHASE OFFSET}

The complex-number DSP output of a Delayen-style SEL controller
should have almost constant real part when correcting for cavity microphonics.
In real time, this can be visualized
in an I \& Q distribution which in LCLS-II operations has been nicknamed
the ``Cheeto'' plot.  If the ``Cheeto'' is
tilted, meaning the detuning signal bleeds into the real part,
the phase offset should be adjusted to reestablish orthogonality.
Human operators can't be expected to do that for 296 cavities!
Instead, we have a python-based process that runs every 20~minutes,
looking for that correlation and adjusting the phase offset
to remove it.
That process can only work when a cavity's field control loops are {\bf not} clipping.

\begin{figure}[hbt]
    \centering
    \includegraphics[scale=0.35]{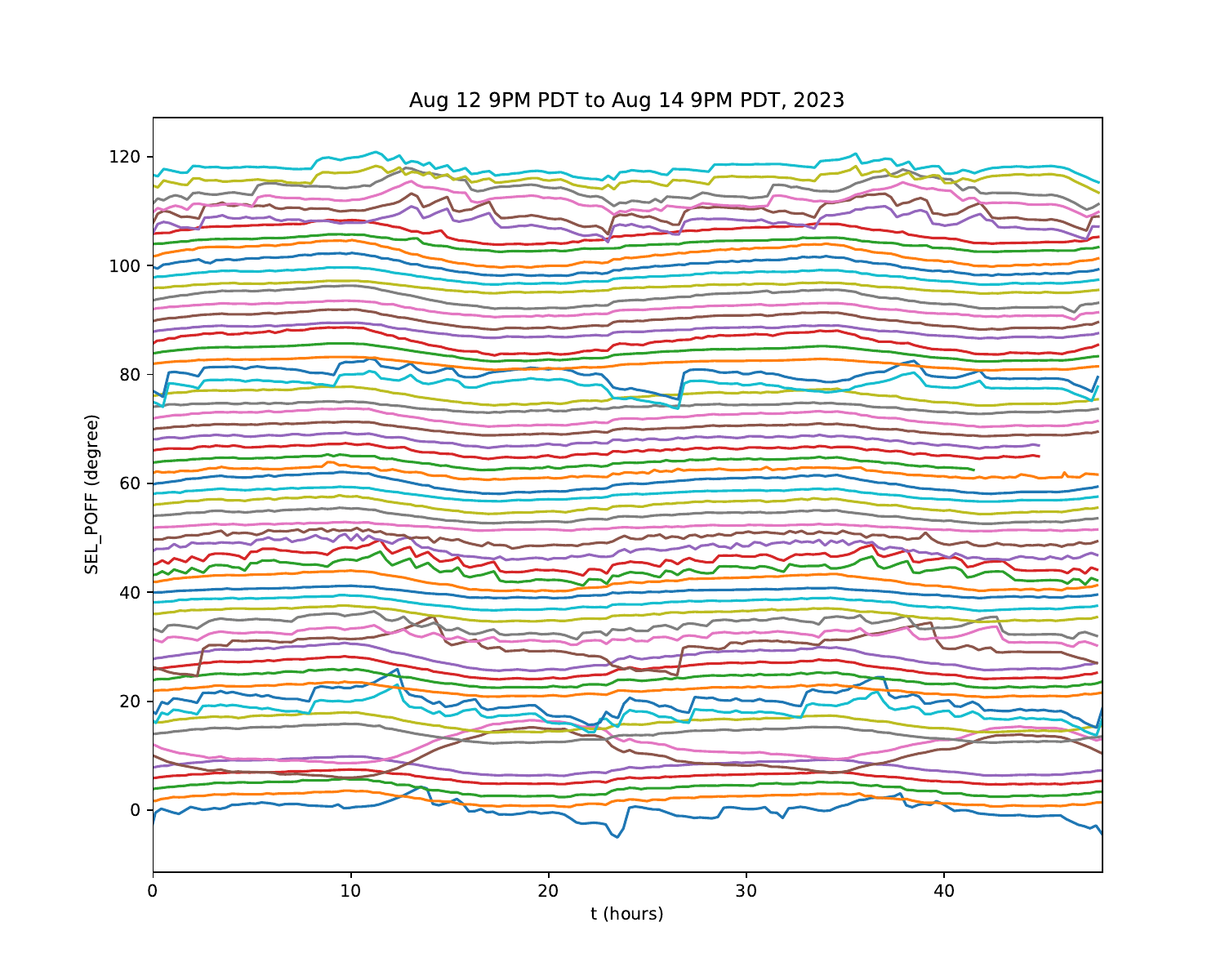}
    \caption{Observed SEL phase offset drift.}
    \label{sel_poff_drift}
\end{figure}

Optimized SEL phase offset changes
observed during normal operations (for a quarter of the cavities)
are shown in figure~\ref{sel_poff_drift}, with a median span of 3.7$^\circ$.

\begin{figure}[hbt]
    \centering
    \includegraphics[scale=0.55]{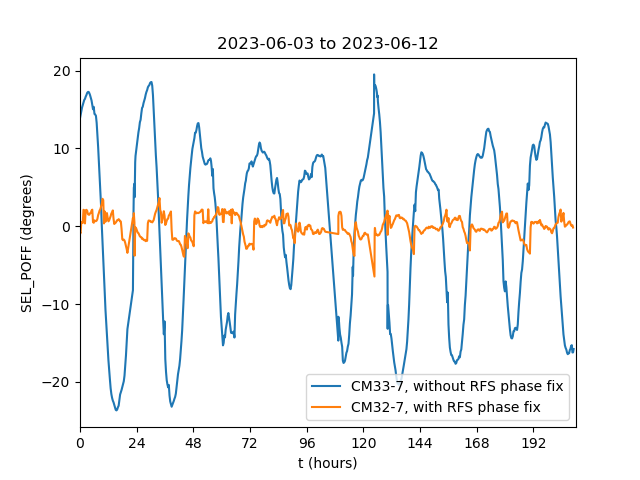}
    \caption{SEL phase offset with and without proper RFS digital phasing.}
    \label{prl_zero_effect}
\end{figure}

System drifts like this got much worse when the PRL phase locking feature
was first added, because it only changed the phase in the PRC, not RFS.
After copying the DDS phase offset from PRC to RFS,
all 18 RF channels (in three chassis) stay phase-stable in the
presence of LO phase drift.
Testing with adjacent cryomodules in L3 shows that the daily span of {\tt SEL\_POFF} was
reduced from 43$^\circ$ to 10$^\circ$, as shown in figure~\ref{prl_zero_effect}.

\section{DETUNING COMPUTATIONS}

Forward and cavity signals are analyzed in terms of
the (complex number) state-space equation
$$ \frac{\mathrm{d}V}{\mathrm{d}t} = aV + bK ~~~. $$
Given $b$, and live measurements of cavity voltage $V$ and drive signal $K$,
the FPGA DSP solves for $a$ every 11.2\thinspace$\mu$s.
This computation works independent of cavity operation mode,
even when (somewhat) off-resonance. It does require non-negligible
energy in the cavity field.

The $b$ coefficient is characterized in the cavity turn-on process,
but will drift with cable electrical length.
Self-tuning of $\angle b$ (or {\tt BCOEFP}) has key similarities to and
differences from the {\tt SEL\_POFF} adjustment process.
This process also depends on properites of the cavity changing
due to microphonics, and also requires the cavity field to be
approximately constant (closed loop).
Without the SSA as part of the characterized plant, however,
standard 2-D statistics can directly find the axes of strong
and weak variation in the measured forward wave signal (in $\sqrt{\rm W}$);
we use numpy covariance (np.cov) and eigenvalue (np.linalg.eig) routines.
This linear-fit analysis does not need an initial guess for $\angle b$.

\begin{figure}[hbt]
    \centering
    \includegraphics[scale=0.35]{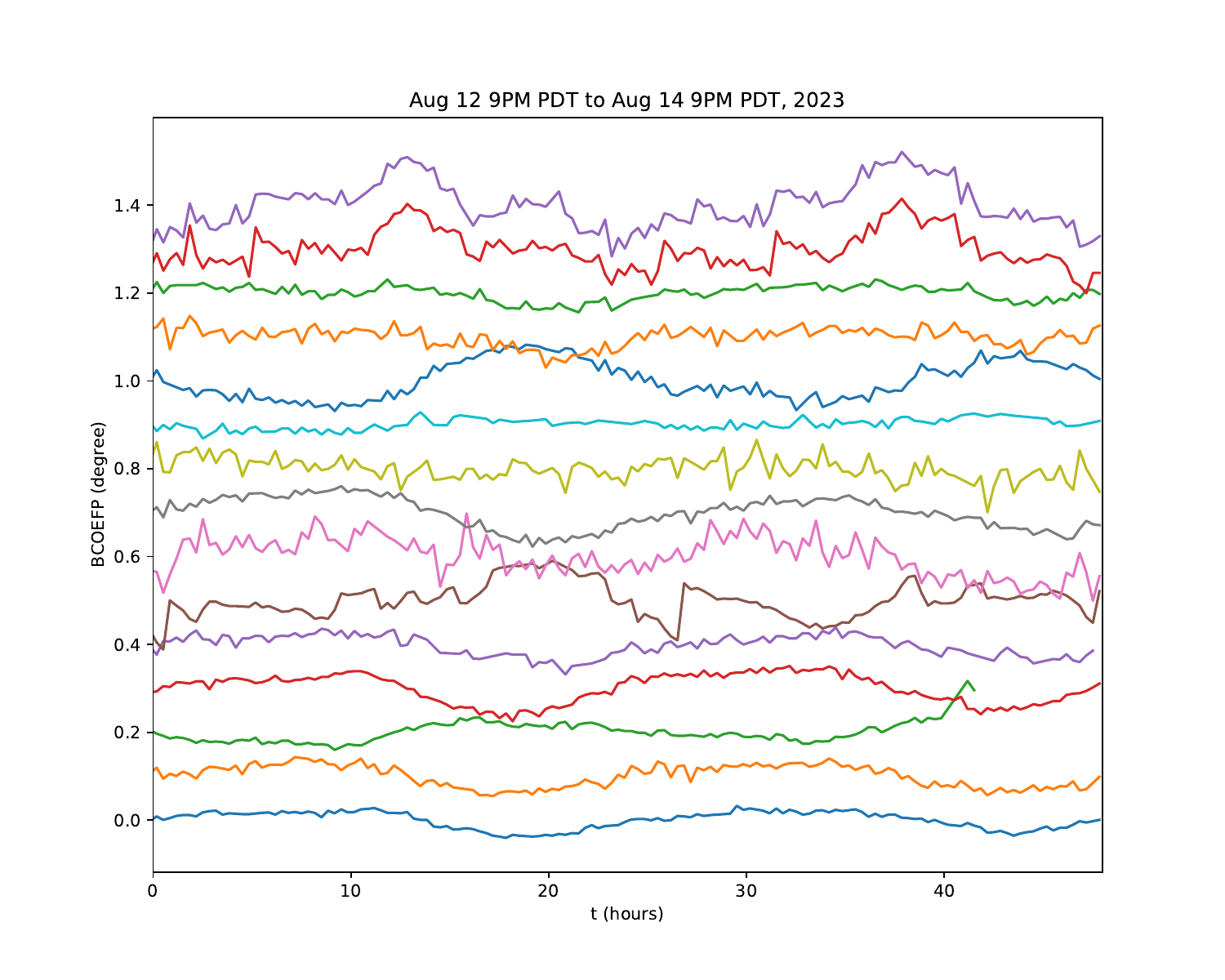}
    \caption{Observed drift of $\angle b$.}
    \label{bcoefp_drift}
\end{figure}

Tests of this process show a median day-night span of 0.14$^\circ$, shown in figure~\ref{bcoefp_drift}.
This is an indication that the forward and cavity cables
to the tunnel are nicely protected from day-night temperature swings,
as originally planned.

\section{BEAM ENERGY NOISE}

\begin{figure}[hbt]
    \centering
    \includegraphics[scale=0.56]{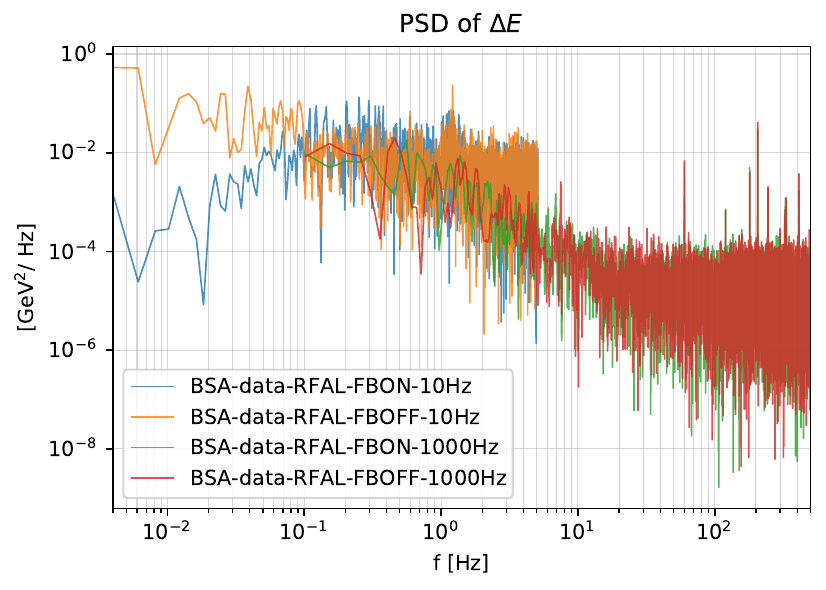}
    \caption{Noise spectrum of Linac energy.}
    \label{psd_plots_compare}
\end{figure}

Preliminary beam energy noise spectra as shown in figure~\ref{psd_plots_compare} have been determined by BPM data capture ~\cite{bsa}, with and without the beam-based energy feedback running.
This represents a first attempt at the out-of-loop characterization
of LCLS-II combined LLRF, PRL, and cavity noise.

Clear signatures of white and $1/f$ noise are seen, along
with the 0.1\thinspace Hz high-pass effect of running feedback; this chart
represents in-loop error for the beam-based feedback.
It's too early to tell what the relative contribution here
is from residual LLRF noise and beam instrumentation noise.

\section{CONCLUSION}

The original (2015) hardware architecture gave enough hooks to
make a usable system: most cavities take care of themselves most of the time.
One can look at the system diagram and identify how electrical-length-drift
of each cable will get self-corrected, except for the key cable-bundle from cavity and PRL to PRC.

Python-based waveform analysis for adjusting {\tt SEL\_POFF} and {\tt BCOEFP} running
every 20~minutes is ``good enough'', so that there hasn't been an urgency to
develop an EPICS-IOC-based {\tt C++} version that could refresh much faster.
More work is yet to come as beam loading is increased and firmware
features are added to support that.

All these features depend on many layers of software and FPGA code.
We have pretty good Continuous Integration (CI) testing,
including hardware-in-the-loop tests, but always wish for more and better.

The system works well, thanks to the hard work of the team.
LCLS-II as a whole reached ``first light'' on Sept.~12, 2023.


\begin{thebibliography}{9} 

\bibitem{lcls2}
  \text{C. Serrano et al.,} ``\textit{Design and Implementation of the LLRF System for LCLS-II},'' ICALEPCS2017, Barcelona, Spain

\bibitem{delayen}
  \text{J. R. Delayen,} ``\textit{Phase and Amplitude Stabilization
of Superconducting Resonators},'' Ph.D. Thesis, 1978

\bibitem{prl1}
  \text{C. Xu et al.,} ``\textit{LCLS-II Phase Reference System},'' LLRF'17, Barcelona, Spain

\bibitem{prl2}
 \text{S.D. Murthy et al.,} ``\textit{Installation, Commissioning and Performance of Phase Reference Line for LCLS-II},'' LLRF'22, Brugg-Windisch, Switzerland

\bibitem{bsa}
 \text{T. Straumann et al.,} ``\textit{FPGA-BASED BPM DATA ACQUISITION FOR LCLS-II,}'' ICALEPCS'17, Barcelona, Spain

\end{thebibliography}
\end{document}